\documentclass[prd,twocolumn,tightenlines,superscriptaddress,nofootinbib]{revtex4}
\usepackage{amssymb,latexsym}
\usepackage{amsmath,amsbsy,bbm}
\usepackage{epsfig,bm}
\usepackage{graphicx,comment}
\usepackage{color}
\usepackage[normalem]{ulem}
\begin{document}

\title{Monte Carlo determination of the critical exponents for a quantum phase 
transition of a dimerized spin-1/2 Heisenberg model}

\author{F.-J. Jiang}
\email[]{fjjiang@ntnu.edu.tw}
\affiliation{Department of Physics, National Taiwan Normal University, 
88, Sec.4, Ting-Chou Rd., Taipei 116, Taiwan}

\vspace{-2cm}

\begin{abstract}
We simulate the spin-1/2 Heisenberg model with a spatially staggered 
anisotropy using first principles Monte Carlo method. In particular,
the critical exponents $\beta/\nu$ and $\omega$ associated with the 
quantum phase transition induced by dimerization are determined with 
high precision. Here $\beta$ and $\nu$ are the exponents related to the 
magnetization and the correlation length, respectively. In addition, 
$\omega$ is the confluent exponent. With very accurate data of the 
relevant observables, we first obtain a value of $\omega$ compatible 
with the known result in the $O(3)$ universality class. Further,  
using either the value of $\omega$ determined here or the established one 
in the literature, the exponent $\beta/\nu$ calculated from our data 
is in quantitative agreement with the known result $\beta/\nu = 0.519(1)$
as well. Our investigation suggests that the quantum phase transition
studied here is fully consistent with the $O(3)$ universality class.

\end{abstract}


\maketitle

\section{Introduction}
Despite their simplicity, Heisenberg-type models continue to be one of the research topics in the 
condensed matter physics. With these models,
one can obtain qualitative, or even quantitative understanding of real materials.
It is also because of this feature of simplicity, various numerical methods
are available to study the properties of Heisenberg-type models with high 
accuracy \cite{White92,White93,Schol05,Bea96,Evertz03,San97,San99,Proko98,Noack05,San10,Gull11,Troyer08,Bau11}.
Indeed, by investigating these models with the various numerical methods, many 
theories, and consequently properties of real materials, are verified and better understood 
\cite{San95,Sac00,Tro02,Melin02,Hog03,Laf03,Lin03,Laf06,Hinkov2007,Hinkov2008,Pardini08,Jiang09.1,Rue08,Kul11,Oti12,Jin12}. 
Hence even those days, these simple models still have attracted a lot of theoretical interests. Among the 
studies of Heisenberg-type models carried out recently, 
one striking observation is the possibility of a new universality class for 
the two-dimensional dimerized spin-1/2 Heisenberg model with a spatially 
staggered anisotropy (2-d staggered-dimer model) \cite{Wenzel08}. It is 
believed that the quantum phase transition induced by dimerization of this model 
should be governed by the $O(3)$ universality class theoretically 
\cite{Chakr88,Haldane88,Chubu94,Sachdev99,Vojta03}. On the other hand, a 
recent large scale Monte Carlo calculation obtains $\nu = 0.689(5)$ and 
$\beta/\nu = 0.545(5)$ which are in contradiction to the established $O(3)$ 
results $\nu = 0.7112(5)$ and $\beta/\nu = 0.519(1)$ in the literature 
\cite{Cam02}. Here $\nu$ and $\beta$ are the critical exponents corresponding 
to the correlation length and the magnetization, respectively. Since this surprising
finding, several efforts have been devoted to study the phase transition 
induced by dimerization of the 2-d
staggered-dimer model. At the moment it is well-established theoretically that because of 
an irrelevant cubic term \cite{Fritz11}, there is a large correction to 
scaling for this phase transition which leads to the unexpected
 $\nu = 0.689(5)$ and $\beta/\nu = 0.545(5)$ obtained in \cite{Wenzel08}. 
Further, Monte Carlo study also provides strong evidence to support the 
scenario of an enhanced correction to scaling for this model\cite{Jiang11.8}.

While theoretically the unexpected results obtained in \cite{Wenzel08} 
can be explained by the cubic term introduced in \cite{Fritz11}, 
the explicit role of the cubic term is not clear at the moment. One natural explanation
for the large correction to scaling due to the cubic term is the reduction
of the magnitude of the confluent exponent $\omega$. Whether this is indeed
the case has not explored yet. Hence in this 
study, we have investigated the phase transition induced by dimerization of 
the 2-d staggered-dimer spin-1/2 Heisenberg model on the square lattice. 
In particular, the numerical value of $\omega$ is determined with high precision.
The exponent $\beta/\nu$ is also calculated as a by production.
From our analysis, we find that our Monte Carlo data 
points are fully compatible with the established results $\omega = 0.78(2)$ 
and $\beta/\nu = 0.519(1)$ in the $O(3)$ universality class. Interestingly, 
while our Monte Carlo data for the considered model are consistent with the 
proposal of an enhanced correction to scaling due to an irrelevant cubic term, 
we find that the irrelevant cubic term likely has little influence on $\omega$. 
Hence the role of the cubic term for the large correction to
scaling observed for the considered quantum phase transition requires
further investigation.

This paper is organized as follows. First, after an introduction, the 
spatially anisotropic quantum Heisenberg model and the relevant observables 
studied in this work are briefly described in section two. Then in section 
three we present our numerical results. In particular, the results obtained 
from the finite-size scaling analysis are discussed in detail. Finally we 
conclude our investigation in section four.

\begin{figure}
\begin{center}
\includegraphics[width=0.28\textwidth]{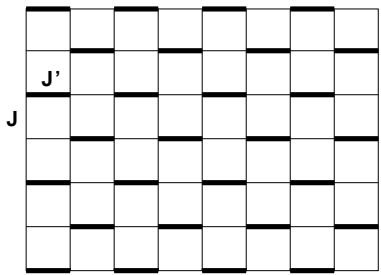}
\end{center}\vskip-0.5cm
\caption{The spatially anisotropic Heisenberg model considered in this study.}
\label{fig1}
\end{figure}

\section{Microscopic Model and Corresponding Observables}
The Heisenberg model considered in this study is defined by the Hamilton 
operator
\begin{eqnarray}
\label{hamilton}
H = \sum_{\langle xy \rangle}J\,\vec S_x \cdot \vec S_{y}
+\sum_{\langle x'y' \rangle}J'\,\vec S_{x'} \cdot \vec S_{y'},
\end{eqnarray}
where $J$ ($J'$) is the antiferromagnetic exchange coupling connecting nearest 
neighbor spins $\langle  xy \rangle$ ($\langle x'y' \rangle$). The 
model described by Eq.~(1) and investigated here is illustrated in fig.~1. To 
study the critical behavior of this model near the transition driven by the 
anisotropy, in particular to determine the confluent exponent $\omega$, 
the second Binder ratio $Q_2$ which is given by
\begin{equation}
Q_2 = \frac{\langle (m_s^z)^2\rangle^2}{\langle (m_s^z)^4\rangle}
,\end{equation}
is calculated in our simulations. Here $m_s^z$ is the $z$ component of the 
staggered magnetization 
$\vec{m}_s = \frac{1}{L^2}\sum_{x}(-1)^{x_1 + x_2}\vec{S}_x$.
Notice the $L$ and $\vec{S}_x$ appearing above are the box sizes used in the calculations and
a spin-1/2 operator at site $x$, respectively.
In additional to $Q_2$, several generalized Binder ratios defined by
\begin{eqnarray}
Q_3 &=& \frac{\langle (m_s^z)^2\rangle^3}{\langle (m_s^z)^6\rangle},\nonumber\\
Q_{31} &=& \frac{\langle (m_s^z)^2\rangle \langle (m_s^z)^4\rangle }{\langle (m_s^z)^6\rangle} \nonumber \\
Q_8 &=& \frac{\langle (m_s^z)^4\rangle^2}{\langle (m_s^z)^2\rangle \langle (m_s^z)^6\rangle},
\end{eqnarray}
are measured in our investigation as well.
By carefully studying the spatial volume dependence 
of these Binder ratios at the critical point $(J'/J)_c$, one can determine $\omega$ 
with high precision.
Similarly, the exponent $\beta/\nu$ is calculated by studying 
the scaling behavior of the observables $\langle |m_s^z| \rangle$ and 
$\langle (m_s^z)^2\rangle$ at $(J'/J)_c$.

\begin{figure}
\label{fig2}
\begin{center}
\vbox{
\includegraphics[width=0.3\textwidth]{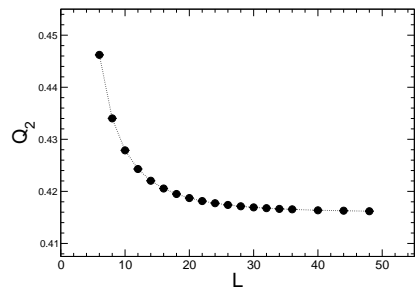}\vskip0.7cm
\includegraphics[width=0.3\textwidth]{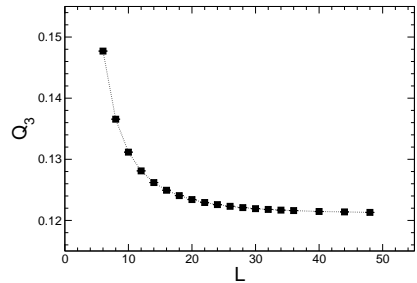}
}
\end{center}
\caption{High accurate $Q_2$ (top panel) and $Q_3$ (bottom panel) data determined at
$(J'/J)_c = 2.51950$ of the 2-d staggered-dimer model.}
\end{figure}

\begin{figure}
\label{fig3}
\begin{center}
\vbox{
\includegraphics[width=0.3\textwidth]{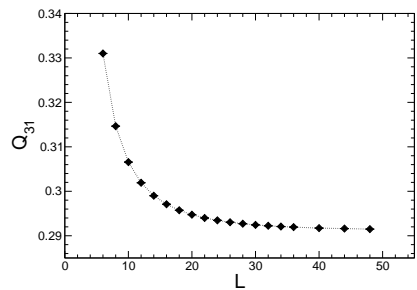}\vskip0.7cm
\includegraphics[width=0.3\textwidth]{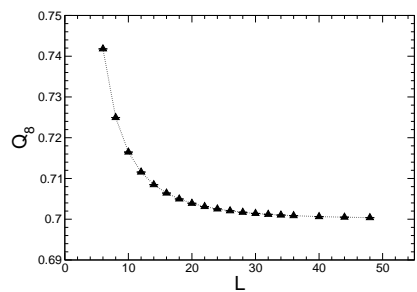}
}
\end{center}
\caption{High accurate $Q_{31}$ (top panel) and $Q_{8}$ (bottom panel) 
data determined at
$(J'/J)_c = 2.51950$ of the 2-d staggered-dimer model.
}
\end{figure}

\section{The Numerical results}

To determine $\beta/\nu$ and $\omega$, we have carried out large scale Monte 
Carlo simulations using the stochastic series expansion algorithm with 
operator-loop update \cite{San99}. Notice the calculations of $\beta/\nu$ and $\omega$ 
require precise knowledge of the critical point $(J'/J)_c$. Specifically, at 
$(J'/J)_c$ the expected finite-size scaling ansatz for 
$\langle |m_{s}^{z}| \rangle$ is given as \cite{Fisher72,Brezin82,Barber83,Brezin85,Fisher89}
\begin{equation}
\langle |m_{s}^{z}| \rangle = (a+bL^{-\omega}+cL^{-2 \omega})L^{-\beta/\nu},
\end{equation}
where $a,\,b,\,c$ are some constants. Similarly, at the critical point,
$\langle (m_{s}^{z})^2 \rangle$ can be described quantitatively by
\begin{equation}
\langle (m_{s}^{z})^2 \rangle = (a'+b'L^{-\omega}+c'L^{-2 \omega})L^{-2\beta/\nu},
\end{equation} 
here $a',b',c'$ are again some constants.
Finally, at $(J'/J)_c$ the 
Binder ratios defined by Eq.~(2) behave like
\begin{equation}
Q_i = (a_i + b_i L^{-\omega} + c_i L^{-2\omega})
\end{equation}  
with some constants $a_i,\,b_i,\,c_i$ as well.

The critical point $(J'/J)_c$ of the considered phase transition has been calculated with high accuracy in 
\cite{Wenzel08,Jiang11.8}. Using the good scaling property of
the observable winding number squared in the 2-direction,
$(J'/J)_c$ is estimated to be $2.51950(3)$ \cite{Jiang11.8}. Hence in our study, we have performed our simulations at 
$(J'/J)_c = 2.51950$ in order to determine the numerical values of 
$\omega$ and $\beta/\nu$. We have additionally carried out some calculations at 
$(J'/J)_c = 2.51953$ and $(J'/J)_c = 2.51947$ so that
the systematic uncertainties for $\omega$ and $\beta/\nu$ due to the error 
of $(J'/J)_c$ are properly taken into account. At $(J'/J)_c = 2.51950$ ($J'/J = 2.51953, 2.51947$), 
the box sizes employed in our calculations range from $L = 6$ to $L = 48$ 
($L = 6$ to $L =40$). We use the relation $\beta J = 4L$ is our simulations as well so
that the validity of the finite-size scaling analysis performed here is guaranteed.
Notice that very high precision related data points
are essential to calculate $\omega$ accurately. Hence, for each $L$ at $J'/J = 2.51947$, $2.51950$, and $2.51953$,
we have carried out at least 20 simulations. In particular, each simulation starts 
with different random seed and contains $2 \times 10^6$ measurements. In 
other words, effectively each data is obtained with at least $4 \times 10^7$ measurements. 
Finally, the estimators for $\langle |m_s^z| \rangle$ and $\langle (m_s^z)^k\rangle$ 
with $k \ge 2$ as described in \cite{San97} are used in our calculations in order to reach a better 
statistic.

\begin{table}
\label{tab1}
\begin{center}
\begin{tabular}{cccc}
\hline
{\text{Observable}} & $L$ & $\omega$ & $\chi^2/{\text{DOF}}$\\
\hline
\hline
$Q_2$ & $6 \le L \le 48 $ & $\,$$\,$0.728(7)  $\,$& 1.2 \\
\hline
$Q_2$ & $6 \le L \le 44$ & $\,$$\,$0.726(8) $\,$& 1.2\\
\hline
$Q_2$  & $6 \le L \le 40$ & $\,$$\,$0.730(9)  $\,$& 0.93 \\
\hline
$Q_2$ & $6 \le L \le 36 $ & $\,$$\,$0.729(10)  $\,$& 1.1 \\
\hline
$Q_2$ & $8 \le L \le 48$ & $\,$$\,$0.718(12) $\,$& 0.93\\
\hline
$Q_2$  & $8 \le L \le 44$ & $\,$$\,$0.710(15)  $\,$& 0.74 \\
\hline
$Q_2$ & $8 \le L \le 40 $ & $\,$$\,$0.715(17)  $\,$& 0.7 \\
\hline
$Q_2$ & $8 \le L \le 36$ & $\,$$\,$0.709(20) $\,$& 0.66\\
\hline
$Q_2$  & $10 \le L \le 48$ & $\,$$\,$0.722(22)  $\,$& 0.98 \\
\hline
$Q_2$  & $10 \le L \le 44$ & $\,$$\,$0.707(27)  $\,$& 0.8 \\
\hline
$Q_3$ & $8 \le L \le 48$ & $\,$$\,$0.775(11) $\,$& 1.05\\
\hline
$Q_3$ & $ 8 \le L \le 44$ & $\,$$\,$0.773(13)$\,$ & 1.1\\
\hline
$Q_3$ & $ 8 \le L \le 40 $ &$\,$ $\,$0.780(16) $\,$ & 0.86 \\
\hline
$Q_3$ & $ 8 \le L \le 36$ & $\,$ $\,$0.780(18) $\,$ & 0.95\\
\hline
$Q_3$  & $10 \le L \le 48$ & $\,$$\,$0.763(19)  $\,$& 0.93 \\
\hline
$Q_3$ & $ 10 \le L \le 44$ &$\,$ $\,$0.754(24) $\,$ & 0.9\\
\hline
\hline
\end{tabular}
\end{center}
\caption{The numerical values of $\omega$ calculated from 
$Q_2$ and $Q_3$ at $(J'/J)_c = 2.51950$ of the 2-d staggered-dimer model. }
\end{table}

\subsection{Determination of the exponent $\omega$}
To determine $\omega$, let us focus on the finite-size scaling analysis of 
the Binder ratios defined by Eqs.~(2) and (3). At the critical point, the 
finite-size scaling ansatz for all the observables $Q_i$ measured in our 
simulations are given by Eq.~(6). Notice in addition to the corrections 
associated with the confluent exponent $\omega$, there are other subleading 
corrections with exponents $\omega' > \omega$. Since the established values 
of $\omega'$ are larger or equal to $2\omega$, it is reasonable to employ the 
scaling ansatz Eq.~(6) for our data analysis. Notice one can define different Binder ratios
using a similar manner. We find
that the ones we define in Eqs.~(2) and (3) have better scaling behavior and receive
less corrections from higher order terms. Using the relevant data points at $(J'/J)_c = 2.51950$,
tables one and two summarize the 
results of the fits associated with the determination of $\omega$. 
We have carried out many fits using data with different range of box sizes in order to understand
how the value of $\omega$ converges with $L$. Notice all the errors quoted in this study
are conservative estimates based on the uncertainties obtained directly from the fits. Although the
numerical values of $\omega$ shown in tables one and two are slightly below the
established $\omega = 0.78(2)$ in the $O(3)$ universality class, they agree 
very well with $\omega = 0.78(2)$ considering the fact that $\omega$ is a subleading exponent. 
Notice the Monte Carlo determination of
$\omega$ presents in \cite{Has01} does not take into account the systematic 
uncertainties due to higher order corrections. In addition, while not being 
investigated systematically, it is interesting to observe that the values of 
$\omega$ determined in \cite{Has01} have a tendency of having smaller magnitude when data 
points of large $L$ are included in the fits. 
Hence one cannot rule out the scenario
that indeed the true numerical value of $\omega$ in the $O(3)$ universality 
class is below $\omega = 0.78(2)$ which is obtained using the series method \cite{Zinn98}. 
Nevertheless, the values of $\omega$ we reach here are in nice agreement with the
well-established result $\omega = 0.78(2)$. Finally to properly take
into account the systematic error of $\omega$ due the the uncertainties of
$(J'/J)_c$, we have performed similar analysis for the data determined at
$J'/J = 2.51947$ and $J'/J = 2.51953$. The related $Q_2$ and $Q_3$ ($Q_{31}$ and $Q_8$)
data points obtained at $J'/J = 2.51947$ ($J'/J = 2.51953$) are presented in 
fig.~4 (fig.~5).  Further, the results of these additional analysis are
shown in tables 3 to 6. The results in tables 3 to 6 indicate that
both the $\omega$ determined at these two values of $J'/J$ are compatible with $\omega = 0.78(2)$
as well. Notice the results associated with $J'/J = 2.51953$ have poor fitting quality compared
to those of $J'/J = 2.51947$ and $J'/J = 2.51950$.
This might due to either the impact from higher order corrections or the value $J'/J = 2.51953$
is slightly away from the true $(J'/J)_c$. Considering the fact that both the fitting quality using
the data determined with $J'/J = 2.51947$ and $J'/J = 2.51950$ are good, it is likely that
$J'/J = 2.51953$ is not consistent with $(J'/J)_c$ when the precision of our data is considered.
Finally, using the results in tables 1 to 6 with absolute error $\le$ 0.025 and $\chi^2/{\text{DOF}} \le 2.0$,
we arrive at $\omega \sim 0.738$. Here we do not quote the error for $\omega$ and it is reasonable to assume
the error of $\omega$ is within 3 to 4 percent.

\begin{table}
\label{tab2}
\begin{center}
\begin{tabular}{cccc}
\hline
{\text{Observable}} & $L$ & $\omega$ & $\chi^2/{\text{DOF}}$\\
\hline
\hline
$Q_{31}$  & $6 \le L \le 48$ & $\,$$\,$0.773(6)  $\,$& 2.8 \\
\hline
$Q_{31}$ & $6 \le L \le 44 $ & $\,$$\,$0.775(7)  $\,$& 2.9 \\
\hline
$Q_{31}$ & $6 \le L \le 40$ & $\,$$\,$0.780(7) $\,$& 2.35\\
\hline
$Q_{31}$  & $6 \le L \le 36$ & $\,$$\,$0.782(8)  $\,$& 2.4 \\
\hline
$Q_{31}$  & $8 \le L \le 48$ & $\,$$\,$0.748(10)  $\,$& 0.83 \\
\hline
$Q_{31}$ & $8 \le L \le 44 $ & $\,$$\,$0.745(12)  $\,$& 0.83 \\
\hline
$Q_{31}$ & $8 \le L \le 40$ & $\,$$\,$0.750(14) $\,$& 0.7\\
\hline
$Q_{31}$  & $8 \le L \le 36$ & $\,$$\,$0.748(16)  $\,$& 0.75 \\
\hline
$Q_{31}$ & $10 \le L \le 48$ & $\,$$\,$0.742(17) $\,$& 0.85\\
\hline
$Q_{31}$  & $10 \le L \le 44$ & $\,$$\,$0.733(22)  $\,$& 0.76 \\
\hline
$Q_{8}$ & $ 6 \le L \le 48$ & $\,$ $\,$0.742(4) $\,$ & 1.7\\
\hline
$Q_{8}$ & $6 \le L \le 44 $ &$\,$ $\,$0.743(5) $\,$ & 1.8 \\
\hline
$Q_{8}$ & $6 \le L \le 40$ & $\,$$\,$0.745(5)$\,$ & 1.6\\
\hline
$Q_{8}$ & $6 \le L \le 36$ & $\,$$\,$0.746(6) $\,$& 1.6\\
\hline
$Q_{8}$  & $8 \le L \le 48$ & $\,$$\,$0.726(8)  $\,$& 0.55 \\
\hline
$Q_{8}$ & $ 8 \le L \le 44$ &$\,$ $\,$0.724(10) $\,$ & 0.55\\
\hline
$Q_{8}$ & $8 \le L \le 40 $ & $\,$$\,$0.726(11)  $\,$& 0.54 \\
\hline
$Q_8$ & $8 \le L \le 36$ & $\,$$\,$0.725(13) $\,$& 0.6\\
\hline
$Q_{8}$ & $10 \le L \le 48 $ & $\,$$\,$0.723(13)  $\,$& 0.57 \\
\hline
$Q_8$ & $10 \le L \le 44$ & $\,$$\,$0.718(17) $\,$& 0.52\\
\hline
$Q_8$ & $10 \le L \le 40$ & $\,$$\,$0.720(19) $\,$& 0.54\\
\hline
\hline
\end{tabular}
\end{center}
\caption{The numerical values of $\omega$ calculated from 
$Q_{31}$ and $Q_{8}$ at $(J'/J)_c = 2.51950$ of the 2-d staggered-dimer model. }
\end{table}

\begin{figure}
\label{fig4}
\begin{center}
\vbox{
\includegraphics[width=0.3\textwidth]{Q2_J2.51947.eps}\vskip0.7cm
\includegraphics[width=0.3\textwidth]{Q3_J2.51947.eps}
}
\end{center}
\caption{High accurate $Q_{2}$ (top panel) and $Q_{3}$ (bottom panel) data determined at
$(J'/J)_c = 2.51947$ of the 2-d staggered-dimer model.
}
\end{figure}

\begin{figure}
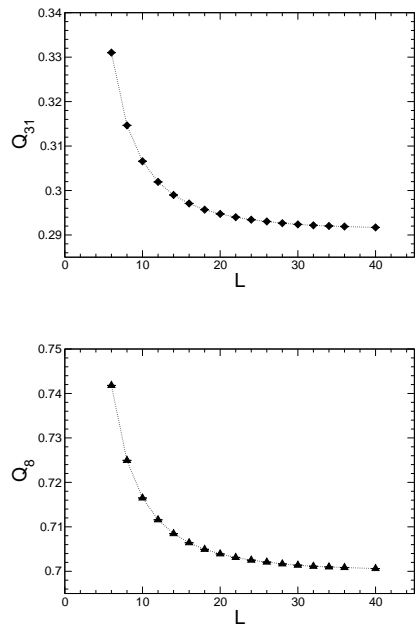

\label{fig5}
\begin{center}
\vbox{
\includegraphics[width=0.3\textwidth]{Q31_J2.51953.eps}\vskip0.7cm
\includegraphics[width=0.3\textwidth]{Q8_J2.51953.eps}
}
\end{center}
\caption{High accurate $Q_{31}$ (top panel) and $Q_{8}$ (bottom panel) data determined at
$(J'/J)_c = 2.51953$ of the 2-d staggered-dimer model.
}
\end{figure}

\subsection{Determination of the exponent $\beta/\nu$}
After having calculated the critical exponent $\omega$ from the relevant 
observables for the quantum phase transition induced by dimerization of the model 
described by fig.~1, we turn to the determination 
of the exponent $\beta/\nu$. To calculate $\beta/\nu$, the scaling behavior of
the observables $\langle | m_s^z | \rangle$ and $\langle (m_s^z)^2 \rangle$ 
are studied. Specifically, at the critical point and for large $L$, the observable $\langle |m_s^z| \rangle$
and $\langle (m_s^z)^2 \rangle$ should scale according to Eqs.~(4) and (5). 
Fig.~6 show our Monte Carlo data of $\langle | m_s^z | \rangle$ and $\langle (m_s^z)^2 \rangle$ 
determined at $(J'/J)_c = 2.51950$. In previous section, we demonstrate that the $\omega$ we obtain 
from several observables are in good agreement with $\omega = 0.78(2)$. Hence, we have fixed $\omega = 0.78$ in our 
analysis of determining $\beta/\nu$. We focus on applying the finite-size scaling
analysis to the relevant data calculated at $(J'/J)_c = 2.51950$.
The obtained values of $\beta/\nu$ are listed in tables 7 and 8. 
From tables 7 and 8, we conclude that the determined $\beta/\nu$ from our data 
with a fixed $\omega = 0.78$ in the fits agree nicely with the established
$\beta/\nu = 0.519(1)$ known in the literature. Notice while the numerical values
of $\omega$ we determine here are in good agreement with $\omega = 0.78(2)$,
they are slightly below $0.78$ with an average $0.738$.  
Interestingly, using a fixed $\omega = 0.738$,
the $\beta/\nu$ determined from the fits are in agreement with 
$\beta/\nu = 0.519(1)$ as well. The results of these new fits using a fixed 
$\omega = 0.738$ are shown in tables 7 and 8 as well. Interestingly, if
we fixed $\beta/\nu$ to be 0.519 in Eqs.~(4) and (5), then the values of $\omega$ obtained
from the fits agree quantitatively with $0.78(2)$. For instance,
applying the ansatz Eq.~(4) (Eq.~(5)) with a fixed $\beta/\nu = 0.519$ to the observable
$\langle | m_s^z | \rangle$ ($\langle (m_s^z)^2 \rangle$) for $8 \le L \le 48$
leads to $\omega = 0.779(9)$ ($\omega = 0.753(7)$) with $\chi^2/{\text{DOF}} \sim 0.8$
($\chi^2/{\text{DOF}} \sim 0.55$). In other words, our data of $\langle | m_s^z | \rangle$ 
and $\langle (m_s^z)^2 \rangle$ are fully compatible with $\beta/\nu = 0.519(1)$
and $\omega = 0.78(2)$ in the $O(3)$ universality class. Notice here the quoted errors
of $\omega$ are the ones directly calculated from the fits.
We have not attempted to performed
a similar detailed analysis for the data obtained at $(J'/J) = 2.51947$ and $(J'/J) = 2.51953$.
Considering the fact that $(J'/J)_c$ is calculated to a very high accuracy, 
it is anticipated that the values of $\beta/\nu$ obtained from the data points
calculated at $J'/J = 2.51947$ and $J'/J = 2.51953$ should be consistent with
$\beta/\nu = 0.519(1)$. We find that indeed this is the case \cite{Jiang13}.
We also observe that the results obtained from the data associated with $J'/J = 2.51947$
have much better $\chi^2/{\text{DOF}}$ than those of $J'/J = 2.51953$.
This provides another evidence that $J'/J = 2.51953$ is slightly 
away from the true critical point.

\begin{table}
\label{tab3}
\begin{center}
\begin{tabular}{ccccc}
\hline
$J'/J$ &{\text{Observable}} & $L$ & $\omega$ & $\chi^2/{\text{DOF}}$\\
\hline
\hline
2.51947 & $Q_2$ & $ 6 \le L \le 40$ &$\,$ $\,$0.717(9) $\,$ & 0.93\\
\hline
2.51947 & $Q_2$ & $6 \le L \le 36 $ & $\,$$\,$0.719(11)  $\,$& 0.98 \\
\hline
2.51947 & $Q_2$ & $6 \le L \le 32$ & $\,$$\,$0.717(12) $\,$& 1.1\\
\hline
2.51947 & $Q_2$  & $8 \le L \le 40$ & $\,$$\,$0.702(18)  $\,$& 0.7 \\
\hline
2.51947 & $Q_2$ & $8 \le L \le 36 $ & $\,$$\,$0.701(22)  $\,$& 0.75 \\
\hline
2.51947 & $Q_2$ & $8 \le L \le 34$ & $\,$$\,$0.693(24) $\,$& 0.59\\
\hline
2.51947 & $Q_3$ & $ 8 \le L \le 40$ & $\,$ $\,$0.768(17) $\,$ & 0.74\\
\hline
2.51947 & $Q_3$ & $ 8 \le L \le 36 $ &$\,$ $\,$0.771(20) $\,$ & 0.77 \\
\hline
2.51947 & $Q_3$ & $8 \le L \le 32$ & $\,$$\,$0.764(24) $\,$& 0.78\\
\hline
2.51947 & $Q_3$  & $10 \le L \le 40$ & $\,$$\,$0.748(32)  $\,$& 0.62 \\
\hline
\hline
\end{tabular}
\end{center}
\caption{The numerical values of $\omega$ calculated from 
$Q_2$ and $Q_3$ at $J'/J = 2.51947$  
of the 2-d staggered-dimer model. }
\end{table}

\begin{table}
\label{tab4}
\begin{center}
\begin{tabular}{ccccc}
\hline
$J'/J$ & {\text{Observable}} & $L$ & $\omega$ & $\chi^2/{\text{DOF}}$\\
\hline
\hline
2.51947 &$Q_{31}$ & $6 \le L \le 40 $ & $\,$$\,$0.772(8)  $\,$& 2.25 \\
\hline
2.51947 &$Q_{31}$ & $6 \le L \le 36$ & $\,$$\,$0.777(9) $\,$& 2.1\\
\hline
2.51947 &$Q_{31}$  & $6 \le L \le 32$ & $\,$$\,$0.779(11)  $\,$& 2.4 \\
\hline
2.51947 &$Q_{31}$ & $8 \le L \le 40 $ & $\,$$\,$0.740(16)  $\,$& 0.65 \\
\hline
2.51947 &$Q_{31}$ & $8 \le L \le 36$ & $\,$$\,$0.742(19) $\,$& 0.7\\
\hline
2.51947 &$Q_{31}$  & $8 \le L \le 32$ & $\,$$\,$0.733(22)  $\,$& 0.62 \\
\hline
2.51947 &$Q_{8}$ & $ 6 \le L \le 40$ & $\,$ $\,$0.741(6) $\,$ & 1.4\\
\hline
2.51947 &$Q_{8}$ & $6 \le L \le 36 $ &$\,$ $\,$0.744(7) $\,$ & 1.3 \\
\hline
2.51947 &$Q_{8}$ & $6 \le L \le 32$ & $\,$$\,$0.744(8)$\,$ & 1.5\\
\hline
2.51947 &$Q_{8}$ & $8 \le L \le 40$ & $\,$$\,$0.722(12) $\,$& 0.59\\
\hline
2.51947 &$Q_{8}$  & $8 \le L \le 36$ & $\,$$\,$0.724(15)  $\,$& 0.62 \\
\hline
2.51947 &$Q_{8}$ & $ 8 \le L \le 32$ &$\,$ $\,$0.715(18) $\,$ & 0.43\\
\hline
\hline
\end{tabular}
\end{center}
\caption{The numerical values of $\omega$ calculated from 
$Q_{31}$ and $Q_{8}$ at $J'/J = 2.51947$ 
of the 2-d staggered-dimer model. }
\end{table}

\begin{table}
\label{tab31}
\begin{center}
\begin{tabular}{ccccc}
\hline
$J'/J$ &{\text{Observable}} & $L$ & $\omega$ & $\chi^2/{\text{DOF}}$\\
\hline
\hline
2.51953 & $Q_2$ & $6 \le L \le 40 $ & $\,$$\,$0.732(10)  $\,$& 1.9 \\
\hline
2.51953 & $Q_2$ & $6 \le L \le 32$ & $\,$$\,$0.745(13) $\,$& 1.6\\
\hline
2.51953 & $Q_2$  & $8 \le L \le 40$ & $\,$$\,$0.711(20)  $\,$& 1.6 \\
\hline
2.51953 & $Q_2$  & $8 \le L \le 36$ & $\,$$\,$0.718(24)  $\,$& 1.6 \\
\hline
2.51953 & $Q_2$  & $8 \le L \le 34$ & $\,$$\,$0.726(27)  $\,$& 1.6 \\
\hline
2.51953 & $Q_2$  & $8 \le L \le 32$ & $\,$$\,$0.733(30)  $\,$& 1.6 \\
\hline
2.51953 & $Q_3$  & $6 \le L \le 40$ & $\,$$\,$0.832(9)  $\,$& 5.0 \\
\hline
2.51953 & $Q_3$ & $6 \le L \le 32 $ & $\,$$\,$0.856(12)  $\,$& 2.6 \\
\hline
2.51953 & $Q_3$ & $8 \le L \le 40$ & $\,$$\,$0.777(18) $\,$& 2.1\\
\hline
2.51953 & $Q_3$ & $8 \le L \le 36$ & $\,$$\,$0.789(22) $\,$& 1.9\\
\hline
2.51953 & $Q_3$  & $8 \le L \le 34$ & $\,$$\,$0.800(25)  $\,$& 1.6 \\
\hline
2.51953 & $Q_3$  & $8 \le L \le 32$ & $\,$$\,$0.809(27)  $\,$& 1.5 \\
\hline
2.51953 & $Q_3$  & $10 \le L \le 40$ & $\,$$\,$0.759(33)  $\,$& 2.2 \\
\hline
\hline
\end{tabular}
\end{center}
\caption{The numerical values of $\omega$ calculated from 
$Q_2$ and $Q_3$ at $J'/J = 2.51953$ 
of the 2-d staggered-dimer model. }
\end{table} 

\begin{table}
\label{tab41}
\begin{center}
\begin{tabular}{ccccc}
\hline
$J'/J$ &{\text{Observable}} & $L$ & $\omega$ & $\chi^2/{\text{DOF}}$\\
\hline
\hline
2.51953 &$Q_{31}$  & $6 \le L \le 40$ & $\,$$\,$0.785(8)  $\,$& 4.0 \\
\hline
2.51953 &$Q_{31}$ & $6 \le L \le 32 $ & $\,$$\,$0.803(11)  $\,$& 2.2 \\
\hline
2.51953 &$Q_{31}$ & $8 \le L \le 40$ & $\,$$\,$0.747(16) $\,$& 2.2\\
\hline
2.51953 &$Q_{31}$  & $8 \le L \le 34$ & $\,$$\,$0.767(22)  $\,$& 1.8 \\
\hline
2.51953 &$Q_{31}$ & $8 \le L \le 32$ & $\,$$\,$0.775(25) $\,$& 1.7\\
\hline
2.51953 &$Q_{31}$  & $10 \le L \le 40$ & $\,$$\,$0.736(30)  $\,$& 2.3 \\
\hline
2.51953 &$Q_{8}$  & $6 \le L \le 40$ & $\,$$\,$0.748(6)  $\,$& 4.7 \\
\hline
2.51953 &$Q_{8}$ & $6 \le L \le 32 $ & $\,$$\,$0.763(8)  $\,$& 2.5 \\
\hline
2.51953 &$Q_{8}$ & $8 \le L \le 40$ & $\,$$\,$0.723(12) $\,$& 3.4\\
\hline
2.51953 &$Q_{8}$  & $8 \le L \le 32$ & $\,$$\,$0.748(17)  $\,$& 2.4 \\
\hline
2.51953 &$Q_{8}$ & $10 \le L \le 40$ & $\,$$\,$0.718(22) $\,$& 3.6\\
\hline
2.51953 &$Q_{8}$  & $10 \le L \le 32$ & $\,$$\,$0.785(40)  $\,$& 2.2 \\
\hline
\hline
\end{tabular}
\end{center}
\caption{The numerical values of $\omega$ calculated from 
$Q_{31}$ and $Q_{8}$ at $J'/J = 2.51953$ 
of the 2-d staggered-dimer model. }
\end{table}

\section{Discussion and Conclusion}
In this study, we investigate the phase transition induced by dimerization for the 
staggered-dimer spin-1/2 Heisenberg model on the square lattice. 
In particular, we determine the values of the exponents $\omega$ and $\beta/\nu$ with 
high accuracy by employing the finite-size scaling analysis to the relevant observables.
We find that both the numerical values of $\omega$ and $\beta/\nu$ determined here 
match very well with the established results $\omega \sim 0.78$
and $\beta/\nu = 0.519(1)$ known in the literature. Our obtained $\omega$ has an average
of $0.738$ which is slightly below $0.78$. Still, the agreement between our result
$\omega \sim 0.738$ and $\omega \sim 0.78$ is reasonably well. Using either $\omega = 0.738$
or $\omega = 0.78$, the numerical values of $\beta/\nu$ determined in this study
agree quantitatively with 
$\beta/\nu = 0.519(1)$. The results reached here and those obtained in
\cite{Jiang11.8} provide convincing evidence that the considered quantum phase transition
is indeed governed by the $O(3)$ universality class.
It is argued in \cite{Fritz11} that the large correction to scaling for the quantum phase transition induced by dimerization of
the staggered-dimer model is due to an irrelevant cubic term. Our study find that the 
obtained $\omega$ is slightly below the predicted $\omega \sim 0.78$. This finding is consistent
with the scenario that the cubic term influences the value of $\omega$. However, since
the difference between our obtained $\omega$ and $\omega = 0.78(2)$ is not significant, whether this is
because of the cubic term needs a careful examination. Further, a slight reduction for the magnitude of $\omega$
cannot fully explain the large correction to scaling observed for the considered quantum phase transition.
Finally, taking into account the fact that the numerical value of $\omega$ has not fully under control \cite{Has01},
it is desirable to carry out a more detailed investigation to examine the role of the cubic term for the quantum phase 
transition considered here. This will require a much precise value of $(J'/J)_c$ than $(J'/J) = 2.51950(3)$ determined in
\cite{Jiang11.8} and is beyond the scope of our study.

\vskip2cm

\begin{figure}
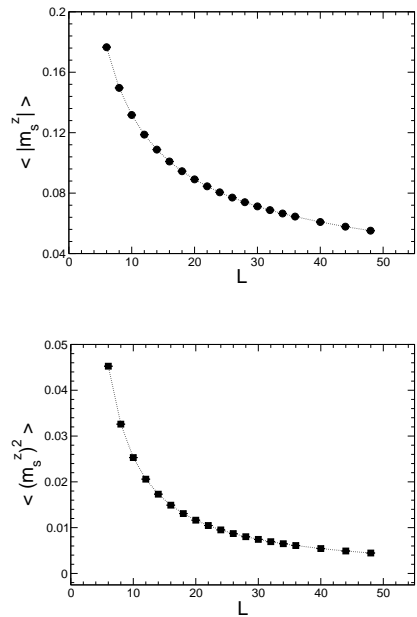

\label{fig6}
\begin{center}
\vbox{
\includegraphics[width=0.3\textwidth]{ms_J2.5195.eps}
\vskip0.7cm
\includegraphics[width=0.3\textwidth]{ms_squared_J2.5195.eps}
}
\end{center}
\caption{High precision data of $\langle |m_s^z|\rangle$ (top panel)
and $\langle (m_s^z)^2\rangle $ (bottom panel), 
obtained at $(J'/J)_c = 2.51950$, of the 2-d staggered-dimer model.}
\end{figure}

\begin{table}
\label{tab5}
\begin{center}
\begin{tabular}{cccc}
\hline
{\text{Observable}} & $L$ & $\beta/\nu$ & $\chi^2/{\text{DOF}}$\\
\hline
\hline
$\langle | m_s^z | \rangle$ & $8 \le L \le 48 $ & 0.517(1) & 0.95\\
\hline
$\langle | m_s^z | \rangle$ & $8 \le L \le 44 $ & 0.5166(12) & 0.81\\
\hline
$\langle | m_s^z | \rangle$  & $8 \le L \le 40 $ & 0.5163(15)  & 0.84 \\
\hline
$\langle | m_s^z | \rangle$  &  $10 \le L \le 48 $ & 0.5184(15)  & 0.6 \\
\hline
$\langle | m_s^z | \rangle$  &  $10 \le L \le 44 $ & 0.5179(18)  & 0.56 \\
\hline
$\langle | m_s^z | \rangle$ & $10 \le L \le 40 $  & 0.5178(23)  & 0.62 \\
\hline
$\langle | m_s^z | \rangle$ & $8 \le L \le 48 $ & 0.5191(9) & 0.8\\
\hline
$\langle | m_s^z | \rangle$ & $8 \le L \le 44 $ & 0.5187(11) & 0.7\\
\hline
$\langle | m_s^z | \rangle$  & $10 \le L \le 48 $ & 0.5200(13)  & 0.56 \\
\hline
$\langle | m_s^z | \rangle$ &  $10 \le L \le 44 $ & 0.5196(16)  & 0.55 \\
\hline
\hline
\end{tabular}
\end{center}
\caption{The numerical values of $\beta/\nu$ calculated from $\langle | m_s^z | \rangle$ 
of the 2-d staggered-dimer model a fixed $\omega = 0.738$ (the top six rows) and a 
fixed $\omega = 0.78$ (the bottom four rows) for the fits.}
\end{table}

\begin{table}
\label{tab6}
\begin{center}
\begin{tabular}{cccc}
\hline
{\text{Observable}} & $L$ & $\beta/\nu$ & $\chi^2/{\text{DOF}}$\\
\hline
\hline
$\langle ( m_s^z )^2 \rangle$ & $8 \le L \le 48 $ & 0.5183(10) & 0.6\\
\hline
$\langle ( m_s^z )^2 \rangle$ & $8 \le L \le 44 $ & 0.5180(12) & 0.55\\
\hline
$\langle ( m_s^z )^2 \rangle$  & $8 \le L \le 40 $ & 0.5178(15)  & 0.58 \\
\hline
$\langle ( m_s^z )^2 \rangle$  &  $10 \le L \le 48 $ & 0.5191(14)  & 0.48 \\
\hline
$\langle ( m_s^z )^2 \rangle$  &  $10 \le L \le 44 $ & 0.5187(18)  & 0.49 \\
\hline
$\langle ( m_s^z )^2 \rangle$ & $10 \le L \le 40 $  & 0.5187(23)  & 0.53 \\
\hline
$\langle ( m_s^z )^2 \rangle$ & $8 \le L \le 48 $ & 0.5204(9) & 0.49\\
\hline
$\langle ( m_s^z )^2 \rangle$ & $8 \le L \le 44 $ & 0.5202(11) & 0.49\\
\hline
$\langle ( m_s^z )^2 \rangle$  & $10 \le L \le 48 $ & 0.5208(13)  & 0.46 \\
\hline
$\langle ( m_s^z )^2 \rangle$ &  $10 \le L \le 44 $ & 0.5206(16)  & 0.48 \\
\hline
\hline
\end{tabular}
\end{center}
\caption{The numerical values of $\beta/\nu$ calculated from $\langle ( m_s^z )^2 \rangle$ 
of the 2-d staggered-dimer model with a fixed $\omega = 0.738$ (the top six rows) and a 
fixed $\omega = 0.78$ (the bottom four rows) for the fits.}
\end{table}

\vskip-3cm
\section{Acknowledgments}
Partial support from NSC and NCTS (North) of R.O.C. is acknowledged.



\end{document}